\documentclass[review, 5p, times, twocolumn]{elsarticle}
\usepackage{lineno,hyperref}
\usepackage{subfigure}
\usepackage{color}
\usepackage{float}
\usepackage{latexsym}
\usepackage{cuted}
\usepackage{xcolor}
\usepackage{mathtools}
\usepackage{amssymb}
\usepackage{balance}
\usepackage{multicol}
\usepackage{lipsum}
\modulolinenumbers[1]

\journal{Phys. Lett. B}

\begin{document}

\begin{frontmatter}

\title{Shear Viscosity of Turbulent Chiral Plasma}
%\tnotetext[mytitlenote]{Fully documented templates are available in the elsarticle package 
%on \href{http://www.ctan.org/tex-archive/macros/latex/contrib/elsarticle}{CTAN}.}

%\author{Avdhesh Kumar, Jitesh R. Bhatt and Predhiman K. Kaw}
%\address{Physical Research Laboratory, Ahmedabad, India}



\author [prl]{Avdhesh Kumar}
\ead {avdhesh@prl.res.in}

\author [prl]{Jitesh R. Bhatt}
\ead {jeet@prl.res.in}

\author [ipr]{Amita Das}
\ead {amita@ipr.res.in}

\author [ipr]{P. K. Kaw}
\ead {kaw@ipr.res.in}
 
\address [prl]{ Physical Research Laboratory, Navrangpura, Ahmedabad 380 009, India.}
\date{\today}
\address [ipr]{Institute for Plasma Research,
Bhat, Gandhinagar 382428, India.}
\begin{abstract}
%This template helps you to create a properly formatted \LaTeX\ manuscript.
It is well known that the difference between the chemical potentials of
left-handed and right-handed particles in a parity violating (chiral) plasma can lead to
an instability.
%It is well known that in a chiral plasma presence of local CP-violation leads to chiral instabilities in quasi-stationary limit. 
We show that the chiral instability may drive turbulent transport.  Further we
estimate the anomalous viscosity of chiral plasma arising
from the enhanced collisionality due to turbulence.%We show, as a 
%result total Shear viscosity of chiral plasma can be
%lowered. 

\end{abstract}

\begin{keyword}
%\texttt{elsarticle.cls}\sep \LaTeX\sep Elsevier \sep template
%\MSC[2010] 00-01\sep  99-00
Chiral Imbalance, Berry curvature, anomalous viscosity
\end{keyword}

\end{frontmatter}

%\linenumbers

\section{Introduction}
The suggestion that the strongly interacting matter created in the relativistic
heavy-ion collision experiments can have local P and CP  violations has
created a lot of excitement. According to  Refs. \cite{Kharzeev98,Kharzeev06,
	Kharzeev08,Kharzeev09} the proposed P and CP violations in QCD can be due to finite nonzero
topological charges present at high-temperature and density. In presence of a very strong
magnetic field (which can be created during the heavy-ion collision)
the nonzero topological charge can induce a net chiral imbalance. 
As a result particles with positive and negative charges will traverse in opposite
directions along the magnetic field and thus a net charge separation can occur. 
This phenomenon is called `chiral magnetic effect' (CME)\cite{Kharzeev08, Fukushima08, Fukushima10}. 
In a different context, this phenomenon
has also been considered in the field of cosmology \cite{Vilenkin, Vachaspati, Arun15}.
Recently an experiment with the STAR detector at RHIC has been performed to observe the CME
by measuring the three particle azimuthal correlators sensitive to the charge separation. 
It has been found that in case of Au-Au and Cu-Cu collisions at $\sqrt s=200$ GeV correlation
of opposite charges separates out \cite{Abelev1, Abelev2} which can be an indication of CME
or P and CP violation. These developments have created a lot of interests in this field.
Theoretical models that study these aspects of strongly interacting matter consider
a plasma of massless fermions which interact with each other in chiral invariant way.
There exists both hydrodynamical and kinetic theory based models  describing 
such a plasma in which the quantum mechanical nature of the chiral anomaly can have
a macroscopic consequences. In this paper we shall focus on the kinetic theory
approach. Recently it was shown that the CME and other CP violating effects 
can be incorporated within a kinetic theory framework
\cite{Son2012, Zahed, Stephanov, Son:2012zy} by using the Berry curvature \cite{Berry}
corrections. The kinetic theory approach is more general in comparison with a hydrodynamical
framework and can be applied to study various equilibrium and nonequilibrium situations.

It should be noted here that the effect of parity violation because of weak-interaction
considered to be important in the context of core collapsing supernova and the formation
of neutron stars\cite{James:1991, Gusakov:2004mj} e.g. the
peculiar velocity of pulsar \cite{Kusenko:2004mm} or in the
generation of magnetic field during the core collapsing
neutron star \cite{Dvornikov:2013bca, Akira:2014,Yamamoto:2015gzz}.
However, the role of parity violating effects due to the strong sector in a quark star is
not fully explored. In the present work we consider the chiral-plasma instability (CPI)
which may arise either in core collapsing supernova due to weak process\cite{Ohnishi:2014uea}
or in a quark matter in the interior of a neutron star due to strong process.
Such instabilities have been studied in the context of electromagnetic
and quark-gluon plasma at finite temperature using the Berry-curvature modified 
kinetic equation\cite{Akamatsu:2013,Kumar:2016xuh}. A similar kind of instability
can exist in  case of a electroweak plasma and early universe \cite{Akira:2014}.
In Ref. \cite{Akamatsu:2013} it was argued that the chiral-imbalance instability can lead to the 
growth of Chern-Simons number (or magnetic-helicity in
plasma physics context) at expense of the chiral imbalance. 
Subsequently in 
Refs.\cite{Akira:2014, Dvornikov:2015}  
it was shown that the generation of magnetic helicity in presence of chiral instability 
may lead to a huge magnetic field 
of the order of $10^{16}$ G in core of a compact star.
Such kind of instabilities was mentioned in Refs. 
\cite{Redlich:1985, Rubakov:1985,Joyce:1997,Liane:2005, Boyarsky:2012} 
in different context and may be seen in heavy ion collisions.
%Note that the CPI \cite{Akamatsu:2013} occurs at
%$1/k\sim 1/(\alpha_s\mu_5)$, where $\alpha_s$ and $\mu_5$
%respectively denote the strong coupling constant and the chiral chemical potential. 
%The hydrodynamic length scale describes particle collisions is $1/(\alpha_s^4 \mu)$
%where $\mu$ is the chemical potential associated with $U(1)$ particle. Further there can be
%a length scales associated with electric and magnetic screening respectively given
%by $1/(\alpha_s\mu)$ and $1/(\alpha_s^2T)$. The collective effects can be very important
%in the regime between the length scales defined by scales $1/\mu$ and $1/(\alpha_s^2\mu)$.
%The length scale for the chiral instability is between these two scales. 
%and it has been
%suggested that some collective effects can reduce the viscosity calculated using  binary
%collision in the context of QGP produced in the heavy-ion 
%collisions \cite{Asakawa:06, Asakawa06, Hong:2014}.

In this paper we calculate the coefficient of shear viscosity arising due to the
CPI generated turbulence transport in a chiral plasma.
By definition, $\eta$ measures ratio of stress to velocity gradient. Stress
in a medium arises because of momentum-transfer/diffusion generated by a velocity
gradient\cite{reif2009fundamentals}. The momentum transfer in a medium is usually governed by collision.
But in case of turbulence interaction between particles and
field can enhance the decorrelation frequency and the effective viscosity can be written as,
\begin{equation}
\eta\sim\frac{Stress}{\nu_{collision}+\nu_{decorrelation}},
\end{equation} 
where, $\nu_{collision}$ and  $\nu_{decorrelation}$
respectively denote the collision and decorrelation
frequencies.  
In the 
case of a neutron star  collision frequency can become
very small as temperature $T$ become small \cite{Chen:2012jc} and thus the decorrelation frequency
can have dominant contribution in determining $\eta$. 

\section{Linear Response Analysis and Chiral Instability}

We start with the Berry-curvature modified collisionless kinetic (Vlasov) equation at the leading order in $A^{\mu}$ 
\cite{Son:2012zy} given as:
\begin{equation}
(\partial_{t}+{\mathbf{v}}\cdot{\mathbf{\partial_{x}}})n_{\mathbf{p}}+(e{\mathbf{E}}+e{\mathbf{v}}\times{\mathbf{B}}-
{{\mathbf{\partial}}_{\mathbf{x}}}\mathbf{\epsilon_{p}})\cdot{\mathbf{\partial}}_{\mathbf{p}}n_{\mathbf{p}}=0
\label{kineqwbcc}
\end{equation}
where $\mathbf{v}=\frac{\bf{p}}{p}$,   
%where $\tilde {\bf v} = \d{\epsilon_{\bf p}}/\d {\bf p}$, $e{\tilde {\bf E}}=e{\bf E} - \d{\epsilon_{\bf p}}/\d {\bf x}$, 
$\mathbf{\epsilon_{p}}=p(1-e{\mathbf{B\cdot\Omega_{p}}})$ and 
${\mathbf{\Omega_{p}}}=\pm{\mathbf{p}}/{2 p^{3}}$ is the
Berry curvature. $\pm$ sign corresponds to right and
left-handed 
fermions respectively. Note that when $\Omega_{\bf p}$=0, 
energy of 
a chiral fermion $\epsilon_{\bf{p}}$ is independent of x,   
Eq.(\ref{kineqwbcc}) reduces to the standard Vlasov equation. 

\noindent

Current density $\bf j$ is defined as:

\begin{equation}
\label{eq:current}
{\bf j} = -e\int \frac{d^3p}{(2\pi)^3}
\Big[\epsilon_{\bf p}{\mathbf{\partial}}_{\mathbf{p}}n_{\mathbf{p}}
+ e\left({\bf \Omega}_{\bf p} \cdot {\mathbf{\partial}}_{\mathbf{p}}n_{\mathbf{p}}\right) \epsilon_{\bf p} {\bf B}
\\ + \epsilon_{\bf p} {\bf \Omega}_{\bf p} \times {\mathbf{\partial}}_{\mathbf{x}}n_{\mathbf{p}} \Big]
+e {\bf E} \times {\bf \sigma},
\end{equation}
 where, ${\mathbf{\partial}}_{\mathbf{P}}=\frac{\partial}{\partial {\bf p}}$ and 
${\mathbf{\partial}}_{\mathbf{x}}=\frac{\partial}{\partial {\bf x}}$.
The $e {\bf E} \times {\bf \sigma}$ of the above equation represents the anomalous Hall current. 
Where $\sigma$ is as follows: 
\begin{equation}
%\label{eq:sigma}
{\bf \sigma} = e\int \frac{d^3p}{(2\pi)^3} {\bf \Omega}_{\bf p} n_{\bf p}.
\label{eq:sigma}
\end{equation}

Let us first consider right handed fermions with chemical potential $\mu_{R}$. In this case we can take equilibrium 
distribution function of the form $n^{0}_{\mathbf{p}}=1/[e^{(\mathbf{\epsilon_{p}}-\mu_{R})/T}+1]$.

Now for a linear response analysis we express Eq.(\ref{kineqwbcc}) and Eq.(\ref{eq:current}) 
by a linear-order deviation in the gauge field. We consider the power counting scheme \cite{Son:2012zy} for gauge field $A_\mu=O(\epsilon)$
and derivatives $O(\delta)$, where $\epsilon$  and $\delta$ are small and independent parameters. In this scheme one can write 
the distribution function in Eq.(\ref{kineqwbcc}) as follows,
\begin{equation}
n_{\mathbf{p}}=n^{0}_{\mathbf{p}}+e (n^{(\epsilon)}_{\mathbf{p}}+n^{(\epsilon\delta)}_{\mathbf{p}}),
\label{expansion of distribution function in terms of `epsilon' and `epsilondelta'}
\end{equation}
where, $n^{0}_{\mathbf{p}}=n^{0(0)}_{\mathbf{p}}+e n^{0(\epsilon\delta)}_{\mathbf{p}}$
with $n^{0(0)}_{\mathbf{p}}=\frac{1}{[e^{(p-\mu_{R})/T}+1]}$
and $n^{0(\epsilon\delta)}_{\mathbf{p}}=\left(\frac{{\mathbf{B}\cdot\mathbf{v}}}{2 p T}\right)
\frac{e^{(p-\mu_{R})/T}}{[e^{(p-\mu_{R})/T}+1]^2}$.

Now from Eq.(\ref{eq:current}), the anomalous Hall-current term $e {\bf E} \times {\bf \sigma}$, 
can be of order $O(\epsilon\delta)$ or higher.  Here 
we are interested in finding  deviations in current up to order $O (\epsilon\delta)$ therefore, 
only $n^{0(0)}_{\mathbf{p}}$ should contribute to ${\bf \sigma}$ in the anomalous Hall term. 
Hence ${\bf \sigma}$ from Eq.(\ref{eq:sigma}) will be 
\begin{equation}
{\bf \sigma}=\frac{e}{2}\int d\Omega d{p}
\frac{\mathbf v}{(1+e^{({p}-\mu_{R})/T})}=0.\label{sigma1}
\end{equation}  
\noindent
 Thus the anomalous Hall current 
term will not contribute.  

Now the kinetic equation (\ref{kineqwbcc})   
at $O(\epsilon)$ and  $O(\epsilon\delta)$ scales of distribution function can be written as,
\begin{equation}
(\partial_{t}+{\mathbf{v}}\cdot{\mathbf{\partial_{x}}})n^{(\epsilon)}_{\mathbf{p}}=-({\mathbf{E}}+
{\mathbf{v}}\times{\mathbf{B}})\cdot{\mathbf{\partial}}_{\mathbf{p}}n^{0(0)}_{\mathbf{p}}
\label{Eqnfrdistbtnfnaodrepsilon}
\end{equation}
\begin{equation}
(\partial_{t}+{\mathbf{v}}\cdot{\mathbf{\partial_{x}}})(n^{0(\epsilon\delta)}_{\mathbf{p}}+n^{(\epsilon\delta)}_{\mathbf{p}})=
-\frac{1}{e}{{{\mathbf{\partial}}_{\mathbf{x}}}\mathbf{\epsilon_{p}}}\cdot{\mathbf{\partial}}_{\mathbf{p}}{n^{0(0)}_{\mathbf{p}}}
\label{kEqnfrdistbtnfnaodrepsilondelta}
\end{equation}
Similarly equation for the current defined in Eq.(\ref{eq:current}) 
at $O(\epsilon)$ and $O(\epsilon\delta)$ can be written as, 
\begin{equation}
 \mathbf{j^{\mu(\epsilon)}}=e^2\int \frac{d^{3}p}{(2\pi)^3}v^{\mu}n^{(\epsilon)}_{\mathbf{p}}
\label{curntwithbccioepsilon}
\end{equation}
\begin{equation}
 \mathbf{j^{i(\epsilon\delta)}}=e^2\int \frac{d^{3}p}{(2\pi)^3}\left[v^{i}n^{(\epsilon\delta)}_{\mathbf{p}}-
 \left(\frac{{v^{j}}}{2 p}\frac{\partial{n^{0(0)}_{\mathbf{p}}}}{\partial{p^{j}}}\right){{B^{i}}}
 -\epsilon^{ijk}\frac{v^{j}}{2 p}
  \frac{\partial{n^{(\epsilon)}_{\mathbf{p}}}}{\partial{x^{k}}}\right]
  \label{curntwithbccioepsilondelta}
\end{equation}
%
%After adding the contribution from all type of species i.e. right/left fermions with charge 
%$e$ and chemical potential $\mu_{R}/\mu_{L}$ as well as right/left handed antifermions with 
%charge $-e$ and chemical potential $-\mu_{R}/\mu_{L}$,

Using Eqs.(\ref{Eqnfrdistbtnfnaodrepsilon}, \ref{kEqnfrdistbtnfnaodrepsilondelta}, 
\ref{curntwithbccioepsilon}, \ref{curntwithbccioepsilondelta}) and the expression $j^{\mu}_{ind}=\Pi^{\mu\nu}(K)A_{\nu}(K)$ 
one can obtain the expression for the spatial part of self energy,   
$\Pi^{ij}=\Pi^{ij}_{+}+\Pi^{ij}_{-}$ for right handed particles. If we have contribution from all type of 
species i.e. right/left fermions with charge 
$e$ and chemical potential $\mu_{R}/\mu_{L}$ as well as right/left handed antifermions with 
charge $-e$ and chemical potential $-\mu_{R}/\mu_{L}$,
then, $\Pi^{ij}_{+}$ (parity even part of polarization tensor) and 
$\Pi^{ij}_{-}$ (parity-odd part) can be written as, 
\begin{equation}
 {\Pi^{ij}_{+}(K)}=m^{2}_{D}\int \frac{d\Omega}{4 \pi}{{v^{i}}{v^{l}}}
 \left(\delta^{jl}+\frac{v^{j}k^{l}}{{\bf{v\cdot k}}+i\epsilon}\right),
\label{selfenergyoepsilon1}
\end{equation}
\begin{eqnarray}
 {\Pi^{im}_{-}(K)}=C_{E}\int 
 \frac{d\Omega}{4 \pi}\Bigg[i{\epsilon}^{iml}k^{l}+i\omega\frac{({\epsilon}^{jlm}v^{i}-{\epsilon}^{ijl}v^{m})k^{l}v^{j}}
 {({\bf{v\cdot k}}+i{\epsilon})}
 \Bigg] 
\label{selfenergyoepsilondelta1}
%\end{align}
\end{eqnarray}
where, 
\begin{eqnarray}
 m^{2}_{D}=-\frac{e^2}{2\pi^{2}}\int_{0}^{\infty}d{p}{{p}}^{2}
 \Bigg[\frac{\partial{n^{0(0)}_{\mathbf{{p}}}}({p}-\mu_{R})}{\partial{{p}}}+
 \frac{\partial{n^{0(0)}_{\mathbf{p}}}({p}+\mu_{R})}{\partial{{p}}}\nonumber\\+
 \frac{\partial{n^{0(0)}_{\mathbf{{p}}}}({p}-\mu_{L})}{\partial{{p}}}+
 \frac{\partial{n^{0(0)}_{\mathbf{{p}}}}({p}+\mu_{L})}{\partial{{p}}}\Bigg]\nonumber\\
 C_{E}=-\frac{e^2}{4\pi^{2}}\int_{0}^{\infty}d{p}{{p}}
 \Bigg[\frac{\partial{n^{0(0)}_{\mathbf{{p}}}}({p}-\mu_{R})}{\partial{{p}}}-
 \frac{\partial{n^{0(0)}_{\mathbf{{p}}}}({p}+\mu_{R})}{\partial{{p}}}\nonumber\\-
 \frac{\partial{n^{0(0)}_{\mathbf{{p}}}}({p}-\mu_{L})}{\partial{{p}}}+
 \frac{\partial{n^{0(0)}_{\mathbf{{p}}}}({p}+\mu_{L})}{\partial{{p}}}\Bigg]
\label{thermalmass}.
\end{eqnarray}
\noindent
Note that while deriving these expression we have chosen the temporal gauge i.e. $A_0=0$.  
It is easy to perform above integartions and  
get $m^{2}_{D}=e^2\left(\frac{\mu_R^2+\mu_{L}^2}{2\pi^2}+\frac{T^2}{3}\right)$ and $C_E=\frac{e^2 \mu_5}{4\pi^2}$, 
where $\mu_5=\mu_R-\mu_{L}$. From here it is clear that  
that when there is no chiral imbalance $C_{E}=0$ whereas $m^2_{D}\neq0$. Introduction of chemical chemical
potential $\mu_5$ for chiral fermions requires some
clarification. Physically it can be interpreted as  
the imbalance between the right handed and left handed fermion and is arises beacause of   
the topological charge\cite{Fukushima08, Redlich:1985}.

%{\color{red}Physically the chiral 
%chemical potential imply an imbalance between the right handed and left handed fermion. This in turn related to 
%the topological charge\cite{Fukushima08, Redlich:1985}. It should be noted here that due to the axial anomaly 
%chiral chemical potential is not associated with any conserved charge. It can still be regarded as `chemical potential' 
%if its variation is sufficiently slow\cite{Akamatsu:2013}. }

Now Maxwell's equation is 
\begin{equation}
 \partial_{\nu}{F^{\mu\nu}}=J^{\mu}_{ind}+J^{\mu}_{ext}
\end{equation}

Taking the fourier transform and using the expression of the 
induced current $j^{\mu}_{ind}=\Pi^{\mu\nu}(K)A_{\nu}(K)$ and choosing temporal gauge $A_0=0$ as one 
can get, 

\begin{equation}
 [({k^2}-{\omega}^{2})\delta^{ij}-k^{i}k^{j}+\Pi^{ij}(K)]E^{j}=i\omega{j^{i}_{ext}}(k). 
\label{eq1}
\end{equation}
\noindent
%From this one can define
One can define inverse of the propagator as, 
\begin{equation}
	[\Delta^{-1}(K)]^{ij}=({k^2}-{\omega}^{2})\delta^{ij}-k^{i}k^{j}+\Pi^{ij}(K).
\label{invprop}
\end{equation}
Dispersion relation can be obtained by finding the poles of $[\Delta(K)]^{ij}$.  
In order to find the poles of the propagator $\Delta^{ij}$ we 
write $\Pi^{ij}$ in a tensor decomposition. For the current problem we need three independent 
projectors, transverse ${P}^{ij}_{T}=\delta^{ij}-k^{i}k^{j}/{k^{2}}$, longitudinal ${P}^{ij}_{L}=k^{i}k^{j}/{k^{2}}$ 
and a parity odd tensor projector ${P}^{ij}_{A}=i\epsilon^{ijk}{\hat{k}}^k$. 
Thus we write $\Pi^{ij}$ as:
\begin{equation}
 \Pi^{ij}=\Pi_{T}{P}^{ij}_{T}+\Pi_{L}{P}^{ij}_{L}+\Pi_{A}{P}^{ij}_{A}
 \label{seexpan}
\end{equation}
where,
$\Pi_{T}$, $\Pi_{L}$ and $\Pi_{A}$ are some scalar functions of $k$ and $\omega$ and need to be determined.

Following the decomposition of $\Pi^{ij}$, one can also decompose $[\Delta^{-1}(k)]^{ij}$ appearing in Eq.(\ref{invprop}) as
\begin{equation}
 [\Delta^{-1}(K)]^{ij}=C_{T}{P}^{ij}_{T}+C_{L}{P}^{ij}_{L}+C_{A}{P}^{ij}_{A}.
 \label{invpropexpan}
\end{equation}
where coefficient $C$'s are related to the scalar functions
defined in Eq.(\ref{seexpan}) by following equation:
\begin{eqnarray}
\nonumber
 C_{T}=k^{2}-{\omega}^{2}+\Pi_{T}, 
 C_{L}=-{\omega}^{2}+\Pi_{L}, 
 C_{A}=\Pi_{A}. \label{eq2}
\end{eqnarray}
Now using Eq.(\ref{seexpan}) one can write 
 $\Pi_{T}=\frac{1}{2}{P}^{ij}_{T}\Pi^{ij}$, $\Pi_{L}={P}^{ij}_{L}\Pi^{ij}$ and $\Pi_{A}=-\frac{1}{2}{P}^{ij}_{A}\Pi^{ij}$ and then using the 
 Eqs.(\ref{selfenergyoepsilon1}-\ref{selfenergyoepsilondelta1}) for $\Pi^{ij}$ one can obtain, 
\begin{eqnarray}
\Pi_{T}=m^{2}_{D}\frac{\omega^{2}}{2 k^{2}}\left[1+\frac{k^{2}-{\omega}^{2}}{2\omega k}\ln\frac{\omega+k}{\omega-k}\right],\nonumber\\
\Pi_{L}=m^{2}_{D}\frac{\omega^2}{k^2}\left[\frac{\omega}{2 k}\ln\frac{\omega+k}{\omega-k}-1\right],\nonumber\\
\Pi_{A}= k C_{E}\left(1-\frac{{\omega}^2}{k^2}\right)\left[1-\frac{\omega}{2 k}\ln\frac{\omega+k}{\omega-k}\right].
\label{isotropic:eq}
\end{eqnarray}
Now using the fact that a vector and its inverse exists in same space, we can expand $[\Delta(K)]^{ij}$ in 
the tensor projector basis as:
\begin{equation}
[\Delta(K)]^{ij}=a{P}^{ij}_{L}+b{P}^{ij}_{T}+c{P}^{ij}_{A}.
\end{equation}
Now using the relation 
$[\Delta^{-1}(K)]^{ij} [\Delta(K)]^{jl}=\delta^{il}$ one can obtain  
the coefficients $a$, $b$, $c$ in terms of the coefficients $C$'s appearing in Eq.(\ref{invpropexpan}). 
Poles of the $[\Delta^{-1}(K)]^{ij}$ can be obtained by equating denominators of the expressions 
for $a$, $b$, $c$ with zero. In the present case we have same denominator for $b$ and $c$ while it is different for $a$ 
therefore the dispersion relation:
\begin{eqnarray}
\label{dispersionrelation2}
C^{2}_{A}-C^{2}_{T}=0,\\
C_{L}=0.\label{dispersionrelation3}
\end{eqnarray}
Here we would like to note that the dispersion relation given by Eq.(\ref{dispersionrelation3}) gives only 
oscillations and do not have instability therefore, it is not of our interest.
Dispersion relation given by Eq.(\ref{dispersionrelation2}) can be written as:
\begin{equation}
\omega^2=k^2+\Pi_{T}\pm\Pi{A} \label{dispersionrelation21}
\end{equation}
In the quasi-static limit i.e. $\left|{\omega}\right|<<k$,
one can write $\Pi_{T}$ 
$\Pi_{L}$ and $\Pi_{A}$ as:
\begin{eqnarray}
{\Pi_{T}}_{\arrowvert_{\left|{\omega}\right|<<k}}=\left(\mp{i}\frac{\pi}{4}\frac{\omega}{k}\right)m^{2}_{D};\nonumber\\
{\Pi_{L}}_{\arrowvert_{\left|{\omega}\right|<<k}}=O(\omega^2/k^2)+.....\nonumber\\
% m^{2}_{D}\left[\mp{i}\frac{\pi}{2}\frac{\omega}{k}-1\right]\nonumber\\
{\Pi_{A}}_{\arrowvert_{\left|{\omega}\right|<<k}}=
-\frac{\mu_5 k e^2}{4 \pi^2}\left(\mp{i}\frac{\pi}{2}\frac{\omega}{k}-1\right).\label{quasilimitofpit}
\end{eqnarray}
In this limit Eq.(\ref{dispersionrelation21}) with the minus sign will give the dispersion 
relation $\omega=i\rho(k)$ where $\rho(k)$ is given by, 
\begin{equation}
\rho(k)=\left(\frac{4\alpha\mu_5}{\pi^{2}m^2_{D}}\right)k^2\left[1-\frac{\pi k}{\mu_5\alpha_e}\right]
\label{eq5} 
\end{equation}
Here we have used  and defined $\alpha=\frac{e^2}{4\pi}$ as the electromagnetic coupling. 
It is clear from Eq.(\ref{eq5}) that $\omega$ is purely an imaginary number and its real-part is zero i.e. $Re(\omega)=0$. 
Positive  $\rho(k)>0$ implies an instability as $e^{-i(i\rho(k))t}\sim e^{+\rho(k)t}$ due net 
chiral chemical potential $\mu_5$. Thus plasma has exponential instability that can drive turbulence. Instability will be maximum 
at $k_{max}=\frac{2\mu_5\alpha}{3\pi}$.
For simplicity, in the next section we consider 
the case of right handed particles only.
%greater than left handed particles i.e. $\mu_R>>\mu_{L}$ so that $\mu_5=\mu_R$. 
\section{Diffusion via nonlinear particle-wave interaction, decorrelation time}
We shall use Resonance Broadening theory \cite{Diamond09,Dupree1,Dupree2,Bian15,Ishihara1,Rudakov}.
First we consider the case of high density and low
temperature, 
%all the particles are highly degenerate and can be placed %on their Fermi surfaces. 
it can shown  $\mathbf{\epsilon_{p}}=p-e\left(\frac{\mathbf{B_{\omega,k}\cdot v}}{2\mu_R}\right)+O(\frac{1}{\mu^{2}})$\cite{Son:2012zy}. Now consider the distribution function,

\begin{equation}
n_{\mathbf{p}}=n^{0(0)}_{\mathbf{p}}+e n^{1}_{\mathbf{p}{\omega, k}}. 
\label{expansion of distribution function}
\end{equation}
where $\langle n_{\mathbf{p}}\rangle=\langle n^{0(0)}_{\mathbf{p}}\rangle$, $\langle \rangle$ represents 
the spatial averaging. $n^{1}_{\mathbf{p}{\omega, k}}$ is the coherent response to field fluctuations. Taking the spatial averaging Berry curvature modified kinetic Eq.(\ref{kineqwbcc}) can be written as, 
\begin{equation}
\partial_{t}\langle n_{\mathbf{p}}\rangle=-e^2\left\langle\left({\mathbf{E}}_{\omega,k}+
{\mathbf{v}}\times{\mathbf{B}}_{\omega,k}+i\mathbf{k}\left(\frac{\mathbf{B_{\omega,k}\cdot v}}{2\mu_R}\right)
\right)\cdot{\mathbf{\partial}}_{\mathbf{p}}n^{1}_{\mathbf{p}{\omega, k}}\right\rangle
\label{kEqnfrdistbtnfnaodrepsilon}
\end{equation}

In the quasilinear theory trajectories of the particles 
are assumed to be unperturbed irrespective of the presence of fluctuating fields. As a result 
coherent response $n^{1}_{\mathbf{p}{\omega, k}}$ has a peak $~1/(\omega-{\bf{k\cdot v}})$.  
In the resonance broadening theory one considers the perturbed trajectories of the 
particles due to effects of random fields and calculate the approximate coherent response function 
$n^{1}_{\mathbf{p}{\omega, k}}$ as an average over a statistical ensemble or perturbed trajectories.
As a results the peak in the coherent response gets broadened\cite{Diamond09,Bian15}. In the case of resonance broadening theory,  
response function can be written as\cite{Diamond09,Bian15};
\small
{\begin{equation}
 n^{1}_{\mathbf{p},\omega k}=\int^{\infty}_0{dt}e^{-i(\omega-\bf{k\cdot v})t} \langle e^{-i k \delta x(t)}\rangle 
 \left(\mathbf{E}_{\omega,k}+\mathbf{v\times B_{\omega,k}}+i\mathbf{k}\left(\frac{\mathbf{B_{\omega,k}\cdot v}}{2\mu_R}\right)\right)
 \cdot\partial_{\mathbf{p}}\langle n_{\mathbf{p}}\rangle
 \label{np`}
\end{equation}}
%\small
%{\begin{equation}
% n^{1}_{\mathbf{p},\omega k}=e\int^{\infty}_0{dt}e^{-i(\omega t-k x)-i k \delta x}
% \left(\mathbf{E}_{\omega,k}+\mathbf{v\times B_{\omega,k}}+i\mathbf{k}\left(\frac{\mathbf{B_{\omega,k}\cdot v}}{2p}\right)\right)
% \cdot\partial_{\mathbf{p}}\langle n_{\mathbf{p}}\rangle
%\end{equation}}
We take Gaussian probability distribution as,
\begin{equation}
pdf[\delta{p}]=\frac{1}{\sqrt{\pi D t}}e^{-\frac{(\delta{p})^2}{D t}}.
\end{equation}
With the above probability distribution one can get,
\begin{equation}
\langle e^{-i k \delta x(t)}\rangle_{pdf}\approx e^{-\frac{t^3}{t^3_{c}}}.\label{eq1n}
\end{equation}

Here, $t_{c}$ is given by following equation,
\begin{equation}
t^3_{c}=\frac{4 \bar{E}^2_{p}}{k^2 D},\label{decorrelationtime1}
\end{equation}
where, 
$\bar{E}^2_{p}\equiv\frac{\int{d^3\mathbf{p}}E_{p}\langle n_{\mathbf{p}}\rangle}
{\int{d^3\mathbf{p}}\langle n_{\mathbf{p}}\rangle}$.

Substituting Eq.(\ref{eq1n}) in Eq.(\ref{np`}) one gets,
\small
{\begin{equation}
 n^{1}_{\mathbf{p},\omega k}=\int^{\infty}_0{dt} e^{-i(\omega -k\cdot v)t-\frac{t^3}{t^3_{c}}}
 \left(\mathbf{E}_{\omega,k}+\mathbf{v\times B_{\omega,k}}+i\mathbf{k}\left(\frac{\mathbf{B_{\omega,k}\cdot v}}{2\mu_R}\right)\right)
 \cdot\partial_{\mathbf{p}}\langle n_{\mathbf{p}}\rangle \label{coherentresponse}
\end{equation}}
Now,
\begin{equation}
\int^{\infty}_0{dt} e^{-i(\omega -k\cdot v)t-\frac{t^3}{t^3_{c}}}\simeq -\frac{i}{\omega-\mathbf{k\cdot v}+i/{t_c}}.
\end{equation}
Using Eq.(\ref{coherentresponse}) one can write the following Diffusion equation,
\begin{equation}
(\partial_{t}-\partial_{\mathbf{p}}\cdot\mathbf{D(p)}\cdot\partial_{\mathbf{p}})\langle n_{\mathbf{p}}\rangle=0, \label{diffusionequation}
\end{equation}

%\begin{equation}
%\mathbf{D(p)}=-\int^{\infty}_0{dt}e^{-i(\omega t-k x)-i k \delta x}
% \left\langle \mathbf{F_{-\omega,-k}}\mathbf{F_{\omega,k}}\right\rangle
%\end{equation}
where, 
\begin{equation}
\mathbf{D(p)}=-\int{d\omega}{d\bf{k}}\left(\mathbf{F_{-\omega,-k}\frac{i}{\omega-\mathbf{k\cdot v}+i/{t_c}}\mathbf{F_{\omega,k}}}\right)
\end{equation}
and 
\begin{equation}
\mathbf{F_{\omega,k}}=e\left({\mathbf{E_{\omega,k}}}+
{\mathbf{v}}\times{\mathbf{B_{\omega,k}}}+i\mathbf{k}\left(\frac{\mathbf{B_{\omega,k}\cdot v}}{2\mu_R}\right)\right).
\end{equation}

%We take Gaussian probability distribution as,
%\begin{equation}
%pdf[\delta{p}]=\frac{1}{\sqrt{\pi D t}}e^{-\frac{(\delta{p})^2}{D t}}
%\end{equation}

%\begin{equation}
%\langle e^{-i(\omega t-k x)-i k \delta x}\rangle_{pdf}\approx e^{-i(\omega -k\cdot v)t-\frac{t^3}{t^3_{c}}}
%\end{equation}

%Here, $t_{c}$ is given by following equation,
%\begin{equation}
%t^3_{c}=\frac{4 \bar{E}^2_{p}}{k^2 D}.\label{decorrelationtime1}
%\end{equation}
%where, 
%$\bar{E}^2_{p}\equiv\frac{\int{d^3\mathbf{p}}E_{p}\langle n_{\mathbf{p}}\rangle}
%{\int{d^3\mathbf{p}}\langle n_{\mathbf{p}}\rangle}$.
%\begin{equation}
%\int^{\infty}_0{dt} e^{-i(\omega -k\cdot v)t-\frac{t^3}{t^3_{c}}}\simeq -\frac{i}{\omega-\mathbf{k\cdot v}+i/{t_c}}.
%\end{equation}
In this problem we are interested in the studying diffusion only due to color magnetic excitaions. In this case the Diffusion coefficient can be written as,
%{\tiny
%\begin{eqnarray}
%D=e^2\sum_{\omega,k}\left(
%{\mathbf{v}}\times{\mathbf{\delta B_{\omega,k}}}-i\mathbf{k}\left(\frac{\mathbf{\delta B_{\omega,k}\cdot v}}{2p}\right)\right)
%\frac{i}{\omega-\mathbf{k\cdot v}+i/{t_c}}\left(
%{\mathbf{v}}\times{\mathbf{\delta B_{\omega,k}}}-i\mathbf{k}\left(\frac{\mathbf{\delta B_{\omega,k}\cdot v}}{2p}\right)\right)
%\end{eqnarray}}

\begin{equation}
D= i e^2\sum_{\omega,k}\frac{\left({\mathbf{v}}\times{\mathbf{\delta B_{\omega,-k}}}-i\mathbf{k}\left(\frac{\mathbf{\delta B_{-\omega,-k}\cdot v}}{2\mu_R}\right)\right)
\left({\mathbf{v}}\times{\mathbf{\delta B_{\omega,k}}}+i\mathbf{k}\left(\frac{\mathbf{\delta B_{\omega,k}\cdot v}}{2\mu_R}\right)\right)}{{\omega-\mathbf{k\cdot v}+i/{t_c}}}. %	
%D= e^2\sum_{\omega,k}\left({\mathbf{v}}\times{\mathbf{B_{\omega,-k}}}-i\mathbf{k}\left(\frac{\mathbf{\delta B_{-\omega,-k}\cdot v}}{2\mu_R}\right)\right)
%\frac{i}{\omega-\mathbf{k\cdot v}+i/{t_c}}\left({\mathbf{v}}\times{\mathbf{B_{\omega,k}}}++i\mathbf{k}\left(\frac{\mathbf{\delta B_{\omega,k}\cdot v}}{2\mu_R}\right)\right)
\end{equation}
%In the above equation factor 2 is incorporated to take into acount right handed antparticle contribution to the diffusion tensor. 
Now, choosing $\mathbf{k}=k \mathbf{\hat{z}}$, $\mathbf{\delta B_{\omega,k}}=\delta B_{\omega,k} \mathbf{\hat{y}}$. 
and considering 
%motion only in x-direction and 
$\omega=i\gamma$. Then the diffusion coefficient,

\begin{equation}
D=e^2\sum_{\omega,k}\frac{\left({v^2_{z}|\delta B_{\omega,k}|^2}{\bf{\hat{x}\bf{\hat{x}}}}+{v_{x}v_{z}|\delta B_{\omega,k}|^2}{\bf{\hat{x}\bf{\hat{z}}}}+{v^2_{x}|\delta B_{\omega,k}|^2}{\bf{\hat{z}\bf{\hat{z}}}}+\frac{{v^2_y}{k^2|\delta B_{\omega,k}|^2}}{2\mu^2_R}{\bf{\hat{z}\bf{\hat{z}}}}\right)}{(\gamma+1/{t_c}+i{k v_z})}
\end{equation}

For strong turbulence we can use approximation  $(1/{t_c})^2>>({k v_z})^2$\cite{Hong:2014}. In this case, at saturation ($\gamma=0$) the diffusion coefficient can be written as,
%\begin{equation}
%D=e^2\sum_{\omega,k}\frac{1/{t_c}\left({v^2_{z}|\delta B_{\omega,k}|^2}{\bf{\hat{x}\bf{\hat{x}}}}+{v_{x}v_{z}|\delta B_{\omega,k}|^2}{\bf{\hat{x}\bf{\hat{z}}}}+{v^2_{x}|\delta B_{\omega,k}|^2}{\bf{\hat{z}\bf{\hat{z}}}}+\frac{{v^2_y}{k^2|\delta B_{\omega,k}|^2}}{2\mu^2_R}{\bf{\hat{z}\bf{\hat{z}}}}\right)}{(1/{t_c})^2+({k v_z})^2}
%\end{equation}
%Condition  for strong turbulence, $(1/{t_c})^2>>({k v_z})^2$. Therefore by 
%ignoring $({k v_z})^2$ one can obtain,
\begin{equation}
D=e^2\sum_{\omega,k}\frac{\left({v^2_{z}|\delta B_{\omega,k}|^2}{\bf{\hat{x}\bf{\hat{x}}}}+{v_{x}v_{z}|\delta B_{\omega,k}|^2}{\bf{\hat{x}\bf{\hat{z}}}}+{v^2_{x}|\delta B_{\omega,k}|^2}{\bf{\hat{z}\bf{\hat{z}}}}+\frac{{v^2_y}{k^2|\delta B_{\omega,k}|^2}}{2\mu^2_R}{\bf{\hat{z}\bf{\hat{z}}}}\right)}{(1/{t_c})}.\label{diffusioncoefficient}
\end{equation}
%In general there will also be contribution from right-handed
%antiparticles as well as left-handed
%particles/antiparticles. Problem gets greatly simplified
%if we consider the case $\mu_R\gg \mu_L\gg T$. In this
%case we can write $\mu_R\sim\mu_5$. 
Now, taking thermal average of velocities and 
using Eqs.(\ref{decorrelationtime1}, \ref{diffusioncoefficient}) one can get the decorrelation time as,
\begin{equation}
\left(\frac{1}{t_c}\right)^4 \sim \frac{e^2 k^2}{4{\bar{E_p}}^2}\sum_{\omega',k'}\left({v^2_{T}|\delta B_{\omega',k'}|^2}+\frac{{v^2_T}{k'^2|\delta B_{\omega',k'}|^2}}{2\mu^2_R}\right),
\label{decorrealtion}
\end{equation}
%$\bar{E}^2_{p}\equiv\frac{\int{d^3\mathbf{p}}E_{p}\langle n_{\mathbf{p}}\rangle}
%{\int{d^3\mathbf{p}}\langle n_{\mathbf{p}}\rangle}$.
where, $v^2_{T}=\frac{\int{d^3\mathbf{p}}v^2_{x,z}\langle n_{\mathbf{p}}\rangle}
{\int{d^3\mathbf{p}}\langle n_{\mathbf{p}}\rangle}$.
This is the relation between $t_c$ and the intensity of color magnetic excitations.
Now we calculate the decorrelation time by incorporating the non-linear corrections in the self energy due to resonance 
broadening.

Thus due to non-linear wave particle interactions self energy calculated in Eqs.(\ref{selfenergyoepsilon1},\ref{selfenergyoepsilondelta1})  
acquires a corrections as $\omega\rightarrow \omega+i/{t_c}$.

\begin{equation}
 {\Pi^{ij}_{+}(K)}=m^{2}_{D}\int \frac{d\Omega}{4 \pi}{{v^{i}}{v^{l}}}
 \left(\delta^{jl}+\frac{v^{j}k^{l}}{{\bf{v\cdot k}}+i/{t_c}}\right),
\label{selfenergyoepsilon12}
\end{equation}
\begin{eqnarray}
 {\Pi^{im}_{-}(K)}=C_{E}\int 
 \frac{d\Omega}{4 \pi}\Bigg[i{\epsilon}^{iml}k^{l}+i\omega\frac{({\epsilon}^{jlm}v^{i}-{\epsilon}^{ijl}v^{m})k^{l}v^{j}}
 {({\bf{v\cdot k}}+i/{t_c})}
 \Bigg]. 
\label{selfenergyoepsilondelta12}
%\end{align}
\end{eqnarray}
%We start with the dispersion relation,
%\begin{eqnarray}
%{\omega}^2=k^2+\alpha\pm\lambda
%\end{eqnarray}
It is important to note that here we have considered only
right handed particles so in the Eq.\ref{thermalmass},  $m^{2}_{D}$
 and $C_E$ will have contribution from right handed particles only. 
Now, using the similar decomposition of self energy as in case of linear stability analysis one can 
calculate $\Pi_{T}$, and $\Pi_A$ to be of the form,
\begin{eqnarray}
\nonumber
{\Pi_{T}}=-\frac{m^{2}_{D}\left(\omega+\frac{i}{t_c}\right)}{4 k}\left[ln\frac{1-\frac{\omega}{k}+\frac{i}{t_c k}}
{1+\frac{\omega}{k}+\frac{i}{t_c k}}\pm i\pi\right]\\ \nonumber+\frac{m^{2}_{D}\left(\omega+\frac{i}{t_c}\right)}{4 k}
\left(\frac{\omega}{k}+\frac{i}{t_c k}\right)\Bigg[2+\left(\frac{\omega}{k}+\frac{i}{t_c k}\right)
\left(ln\frac{1-\frac{\omega}{k}+\frac{i}{t_c k}}
{1+\frac{\omega}{k}+\frac{i}{t_c k}}\pm i\pi\right)\Bigg],\nonumber
\end{eqnarray}
\begin{eqnarray}
\Pi_{L}=-\frac{m^{2}_{D}\left(\omega+\frac{i}{t_c}\right)}{2 k}
\left(\frac{\omega}{k}+\frac{i}{t_c k}\right)\Bigg[2+\left(\frac{\omega}{k}+\frac{i}{t_c k}\right)
\left(ln\frac{1-\frac{\omega}{k}+\frac{i}{t_c k}}
{1+\frac{\omega}{k}+\frac{i}{t_c k}}\pm i\pi\right)\Bigg],\nonumber
\end{eqnarray}
\begin{eqnarray}
\nonumber
\Pi_{A}= k C_{E}\Bigg[1+\frac{\omega}{2 k}\left(ln\frac{1-\frac{\omega}{k}+\frac{i}{t_c k}}
{1+\frac{\omega}{k}+\frac{i}{t_c k}}\pm i\pi\right)\\-\frac{\omega}{2}\left(\frac{2}{k}\left(\frac{\omega}{k}+\frac{i}{t_c k}\right)+
\frac{1}{k}\left(\frac{\omega}{k}+\frac{i}{t_c k}\right)^2\left(ln\frac{1-\frac{\omega}{k}+\frac{i}{t_c k}}
{1+\frac{\omega}{k}+\frac{i}{t_c k}}\pm i\pi\right)\right)\Bigg].
\label{isotropic:eq}
\end{eqnarray}
Now at saturation $\omega=0$, therefore
{\small{
\begin{eqnarray}
\nonumber
\Pi_{T}=-\frac{i m^{2}_{D}}{4 k t_c}\left[-2i \arctan{\frac{1}{t_c k}}\pm i\pi\right]-\frac{m^{2}_{D}}{4 k^2 t^2_c}\Bigg[2+\frac{i}{t_c k}
\left(-2i\arctan{\frac{1}{t_c k}}\pm i\pi\right)\Bigg]
,\nonumber\\
\Pi_{L}=+\frac{m^{2}_{D}}{4 k^2 t^2_c}\Bigg[2+\frac{i}{t_c k}
\left(-2i \arctan{\frac{1}{t_c k}}\pm i\pi\right)\Bigg],\nonumber\\
\Pi_A= k C_{E}.\nonumber
%\label{isotropic:eqn:omega=0}
\end{eqnarray}}}
We determine the decorrelation time from the dispersion relation given in Eq.(\ref{dispersionrelation21}) with $\omega=0$. 
%\begin{equation}
%k^2-\frac{m^{2}_{D}}{2 k t_c}\left[\arctan{\frac{1}{t_c k}}\mp\frac{\pi}{2}\right]-\frac{m^{2}_{D}}{2 k^2 t^2_c}-
%\frac{m^{2}_{D}}{2 k^3 t^3_c}
%\left(\arctan{\frac{1}{t_c k}}\mp \frac{\pi}{2}\right)\pm k C_{E}=0 \label{decorrelationtime}
%\end{equation}
\begin{equation}
k^2-\frac{m^{2}_{D}}{2 k t_c}\left[\arctan{\frac{1}{t_c k}}-\frac{\pi}{2}\right]-\frac{m^{2}_{D}}{2 k^2 t^2_c}-
\frac{m^{2}_{D}}{2 k^3 t^3_c}
\left(\arctan{\frac{1}{t_c k}}-\frac{\pi}{2}\right)-k C_{E}=0 \label{decorrelationtime}
\end{equation}
This is transcendental equation decorrelation can be obtained by solving this equation.

We consider the case $\mu_R\gg T$, in this case 
$m_D\sim\left(\frac{2\alpha}{\pi}\right)^{1/2}\mu_R$.
Further we consider $k=k_{max}=\frac{2\mu_R\alpha}{3\pi}$
which correspond to maximum growth rates of chiral 
instability. In this case decorrelation time will be dependent
on $\alpha$ and $\mu_R$. For $\alpha=1/137$ the solution for
$1/t_c$ of above equation in terms of $\mu_R$ is shown in
the following figure.        

%Now we have,
%\begin{equation}
%\arctan{x}=\pm\frac{\pi}{2}-\frac{1}{x}+\frac{1}{3 x^3}-\frac{1}{5 x^5}+...........\Big\{^{+if x\geq1}_{-if x\leq-1}\\ \nonumber
%\end{equation}

%\begin{equation}
%\arctan{x}=x-\frac{x^3}{3}+\frac{x^5}{5}+\frac{x^7}{7}...........{if \ |x|<1}\\ \nonumber
%\end{equation}

%Therefore for strong turbulence i.e. when $\frac{1}{t_c k}>1$ Eq.(\ref{decorrelationtime}) will yield,

%\begin{equation}
%t^2_c=\pm\frac{15\mu\alpha}{\pi k m^2_D}\left(1\pm\frac{\pi k}{\mu\alpha}\right)+\frac{5}{k^2}
%\end{equation}

%For weak turbulence we have $\frac{1}{t_c k}<<1$, in this limit Eq.(\ref{decorrelationtime}) will give,

%\begin{equation}
%t_c=\frac{\pi^2 m^2_D}{4\mu\alpha k^2\left(1\pm\frac{\pi k}{\mu\alpha}\right)}
%\end{equation}
\begin{figure}[H]
\begin{center}
\subfigure[]{\includegraphics[bb=0 0 400 270,width=0.40\textwidth]{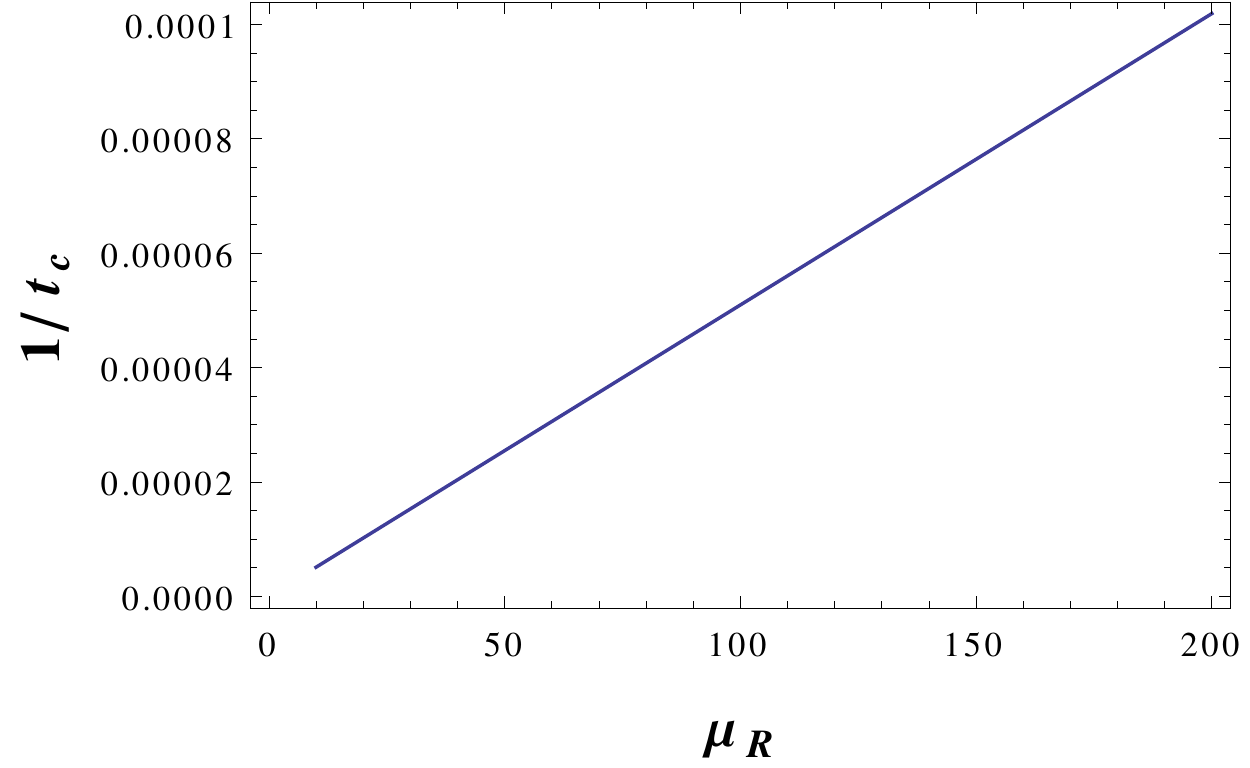}}
%\subfigure[]{\includegraphics[bb=0 0 450 270,width=0.40\textwidth]{fig2.pdf}}
%\subfigure[]{\includegraphics[bb=0 0 400 250,width=0.40\textwidth]{fig3.pdf}}
 \caption 
 {\label{fig[1]} Shows plot of decorrelation frequency $1/t_c$ as a function of chiral chemical potential  $\mu_5$.} 
\label{fig[1]}
\end{center}
\end{figure}
%Now the condition for strong turbulence i.e. $1/{(t_c k)}>>v_z$  can also be written as $1/{(\frac{t_c k}{m_D})}>>m_D v_z$ or 
%$1/{(\frac{t_c k}{m_D})}>>\left(\frac{2\alpha}{\pi}\right)\mu_R$. 
%One can see from above graph that condition 
%for strong turbulence i.e $1/{(\frac{t_c k}{m_D})}>>\left(\frac{2\alpha}{\pi}\right)\mu_R v_z$ satisfied for certain 
%range of $\mu_R$. 
Note that the strong turbulence require that the condition $\frac{1}{t_{c}k_{max}}\gg v_z$
is satisfied in $\mu_R\gg T$ regime. 
%Note that there will be strong turbulence if we have $\frac{1}{t_{c}k_{max}}\gg v_z$. Which indeed a possible scenario in case of electromagnetic chiral plasma or QGP inside the neutron star. 
Now if we take   
$t_c\sim \frac{1}{k}$, $k=k_{max}=\frac{2 \mu_R \alpha}{3\pi}$ and  $\bar{E_p}\sim\mu_R$ we can determine the saturation level of color magnetic excitations using Eq.(\ref{decorrealtion}) as,  
\begin{eqnarray}
\delta B_{\omega,k}\sim \frac{\mu^2_R}{\sqrt{\alpha}}.
\end{eqnarray}

%%%%%%%%%%%%%%%%%%%%%%%%%%%%%%%%%%%%%%%%%%%%%%%%%%%%%%%%%%%%%%%%%%%%%%%%%%%%%%%%%%%%%%%%%%%%%

\section{Calculation of anomalous viscosity}

We follow Ref.\cite{TAbe1,TAbe2} to calculate the anomalous viscosity. For Simplicity we make 
$v_x$ depend on $x$ as,
\begin{equation}
 v_{x}\rightarrow v_{x}-u(x)
\end{equation}
where, $u(x)$ is the mean flow variable.

Now using Eq.(\ref{diffusioncoefficient}) one can write the diffusion equation (Eq.(\ref{diffusionequation})) as,

\begin{eqnarray}
\nonumber
(\partial_{t}+v\cdot \partial_{x})\langle n_{\mathbf{p}}\rangle\simeq
{e^2}\sum_{\omega,k}\frac{1}{1/{t_c}}\Bigg(\left(v^2_{T}|\delta  B_{\omega,k}|^2\right)\partial^2_{\mathbf{p_x}}\langle n_{\mathbf{p}}\rangle+\\ \left({v^2_{T}|\delta B_{\omega,k}|^2}+ \frac{{v^2_T}{k^2|\delta B_{\omega,k}|^2}}{4\mu^2_R}\right)\partial^2_{\mathbf{p_z}}\langle n_{\mathbf{p}}\rangle\Bigg).
\label{diffusionequation1}
\end{eqnarray}

Second term can be written as; 
\begin{equation}
(v\cdot \partial_{x})\langle n_{\mathbf{p}}\rangle\simeq-v^2_{T}p \frac{d\langle n_{\mathbf{p}}\rangle}{dp}\partial_{x}u(x).
\end{equation}

Here, we bring back the term $v\cdot\partial_{x}$. Now, if we consider, $k=k_{max}$, in this case the summation on $\omega$ and $k$ can be lifted 
out and we can write the diffusion equation in terms of mean flow variable as,
\begin{eqnarray}
\nonumber
\partial_{t}\langle n_{\mathbf{p}}\rangle-v^2_{T}p \frac{d\langle n_{\mathbf{p}}\rangle}{dp}\partial_{x}u(x)\simeq
\frac{e^2}{1/{t_c}}\Bigg(\left(v^2_{T}|\delta  B_{\omega,k}|^2\right)\partial^2_{\mathbf{p_x}}\langle n_{\mathbf{p}}\rangle+\\ \left({v^2_{T}|\delta B_{\omega,k}|^2}+ \frac{{v^2_T}{k^2|\delta B_{\omega,k}|^2}}{4\mu^2_R}\right)\partial^2_{\mathbf{p_z}}\langle n_{\mathbf{p}}\rangle\Bigg).
\label{diffusionequation1}
\end{eqnarray}
Note that in the above equation $\omega$ and $k$ respectively corresponds to $\omega_{max}$ and $k_{max}$. 
Now taking moment $\int \frac{d^3p}{(2\pi)^3}\frac{(1+e\delta{\bf{B\cdot\Omega_p}})}{\epsilon_p}(2p^2_x-p^2_y-p^2_z)$, the left 
hand side of above equation will become,
\begin{eqnarray}
\nonumber
LHS=\partial_{t}(2T^{xx}-T^{yy}-T^{zz})-
\Big(\frac{v^2_{T}}{(2\pi)^3}\int d\Omega(2v^2_{x}-v^2_y-v^2_z)\times\\ \frac{(e\delta{\bf{B_{\omega,k}\cdot v}})^2}{4\mu^4_R}\Big)\left[\int^{\infty}_0 
dp p^4\frac{d\langle n_{\mathbf{p}}\rangle}{dp}
\right]\partial_{x}u(x) \label{LHS}
\end{eqnarray}
Note that in the above expression we have used the definition of energy momentum tensor as,

\begin{equation}
T^{\mu\nu}_{\omega,k}=\int \frac{d^3p}{(2\pi)^3}\frac{(1+e\delta{\bf{B_{\omega,k}\cdot\Omega_p}})}
{\epsilon_p}{p^{\mu}p^{\nu}}\langle n_{\mathbf{p}}\rangle.\label{energy-momentum}
%=\frac{1}{(2\pi)^3}\int 
%{d\Omega}\frac{p^{\mu}p^{\nu}}{E}\langle n_{\mathbf{p}}\rangle
\end{equation}

Simplifying above Eq.(\ref{LHS}) we can write,

\begin{eqnarray}
\nonumber
LHS=\partial_{t}(2T^{xx}-T^{yy}-T^{zz})+
\Big(\frac{v^2_{T}}{(2\pi)^3}\int d\Omega(2v^2_{x}-v^2_y-v^2_z)\times\\ \frac{(e\delta{\bf{B_{\omega,k}\cdot v}})^2}{4\mu^4_R}\Big)\left[\int^{\infty}_0 
dp 4p^3{\langle n_{\mathbf{p}}\rangle}
\right]\partial_{x}u(x). 
\end{eqnarray}

%\begin{eqnarray}
%\nonumber
%RHS=\frac{{e^2 v^2_T}{k^2|\delta B_{\omega,k}|^2}}{4\mu^2_R}\frac{1}{1/{t_c}}
%\Bigg(\frac{1}{(2\pi)^3}\int d\Omega(2v^2_{x}-v^2_y-v^2_z)\times\\ \nonumber
%\int^{\infty}_0 dp p^3 \partial^2_{\mathbf{p_z}}\langle n_{\mathbf{p}}\rangle
%+\frac{1}{(2\pi)^3}\int d\Omega(2v^2_{x}-v^2_y-v^2_z)\times\\ \frac{(e\delta{\bf{B_{\omega,k}\cdot v}})^2}{4\mu^3_R}\Big[\int^{\infty}_0 
%dp p^2 \partial^2_{\mathbf{p_z}}\langle n_{\mathbf{p}}\rangle
%\Big]\Bigg).
%\end{eqnarray}
Now,
\begin{eqnarray}
\nonumber
RHS=\frac{{{e^2 v^2_T}{|\delta B_{\omega,k}|^2}}}{1/{t_c}}
\Bigg(\frac{1}{(2\pi)^3}\int d\Omega(2v^2_{x}-v^2_y-v^2_z)\frac{(e\delta{\bf{B_{\omega,k}\cdot v}})^2}{4\mu^4_R}\times\\ \nonumber
\left(\int^{\infty}_0 dp p^3\partial^2_{\mathbf{p_x}}\langle n_{\mathbf{p}}\rangle+\left(1+\frac{k^2}{4\mu^2_R}\right)\int^{\infty}_0 
dp p^3 \partial^2_{\mathbf{p_z}}\langle n_{\mathbf{p}}\rangle
\right)\Bigg).
\end{eqnarray}
Now using,
\begin{equation}
 \partial^2_{\mathbf{p_z}}\langle n_{\mathbf{p}}\rangle=\frac{p^2_z}{p^2}d^2_{p}\langle n_{\mathbf{p}}\rangle+
 \frac{1}{p} d_{p}\langle n_{\mathbf{p}}\rangle-
 \frac{p^2_z}{p^3}d_{p}\langle n_{\mathbf{p}}\rangle.
\end{equation}
and writing $\partial^2_{\mathbf{p_x}}\langle n_{\mathbf{p}}\rangle$ in a similar fashion, One can get;  
%\begin{equation}
%\partial^2_{\mathbf{p_x}}\langle n_{\mathbf{p}}\rangle=\frac{p^2_x}{p^2}d^2_{p}\langle n_{\mathbf{p}}\rangle+
%\frac{1}{p} d_{p}\langle n_{\mathbf{p}}\rangle-
%\frac{p^2_x}{p^3}d_{p}\langle n_{\mathbf{p}}\rangle.
%\end{equation}
\begin{eqnarray}
\nonumber
RHS=\frac{{{e^2 v^2_T}{|\delta B_{\omega,k}|^2}}}{1/{t_c}}
\Bigg(\frac{1}{(2\pi)^3}\int d\Omega(2v^2_{x}-v^2_y-v^2_z)\frac{(e\delta{\bf{B_{\omega,k}\cdot v}})^2}{4\mu^4_R}\times\\ \nonumber
\Bigg(\int^{\infty}_0 dp p^3\left({v^2_x}d^2_{p}\langle n_{\mathbf{p}}\rangle+
\frac{1}{p} d_{p}\langle n_{\mathbf{p}}\rangle-
\frac{v^2_x}{p}d_{p}\langle n_{\mathbf{p}}\rangle\right)+\\ \nonumber\left(1+\frac{k^2}{4\mu^2_R}\right)\int^{\infty}_0 dp p^3 \left({v^2_z}d^2_{p}\langle n_{\mathbf{p}}\rangle+
\frac{1}{p} d_{p}\langle n_{\mathbf{p}}\rangle-
\frac{{v^2_z}}{p}d_{p}\langle n_{\mathbf{p}}\rangle\right)
\Bigg)\Bigg).
\end{eqnarray}
With further simplification we can write above  equation as,
\begin{eqnarray}
\nonumber
RHS=\frac{{{e^2 v^2_T}{|\delta B_{\omega,k}|^2}}}{1/{t_c}}
\Bigg(\frac{1}{(2\pi)^3}\int d\Omega(2v^2_{x}-v^2_y-v^2_z)\frac{(e\delta{\bf{B_{\omega,k}\cdot v}})^2}{4\mu^4_R}\times\\ \nonumber
\Bigg(\int^{\infty}_0 dp\left(8{v^2_x}p\langle n_{\mathbf{p}}\rangle-
2 p\langle n_{\mathbf{p}}\rangle\right)+\\ \nonumber\left(1+\frac{k^2}{4\mu^2_R}\right)\int^{\infty}_0 dp\left(8{v^2_z}p\langle n_{\mathbf{p}}\rangle-
2 p\langle n_{\mathbf{p}}\rangle\right)
\Bigg)\Bigg).
\end{eqnarray}
We choose stationary limit in this case
$\partial_{t}(2T^{xx}-T^{yy}-T^{zz})=0$, therefore from the
diffusion equation (L.H.S=R.H.S) 
we can get,
%\begin{eqnarray}
%\nonumber
%-{v^2_{T}}\int \frac{d\Omega}{(2\pi)^3}\frac{(e\delta{\bf{B_{\omega,k}\cdot v}})^2(2v^2_{x}-v^2_y-v^2_z)}{4}\left[\int^{\infty}_0 
%dp \frac{d\langle n_{\mathbf{p}}\rangle}{dp}
%\right]\partial_{x}u(x)= \\\sum_{\omega,k}\frac{e^2{v^2_T}{k^2|\delta B_{\omega,k}|^2}}{4E^{2}_{T}\frac{1}{1/{t_c}}}
%\int \frac{d\Omega}{(2\pi)^3}\frac{(e\delta{\bf{B_{\omega,k}\cdot v}})^2(2v^2_{x}-v^2_y-v^2_z)}{4}\left[\int^{\infty}_0 
%dp \partial^2_{\mathbf{p_z}}\langle n_{\mathbf{p}}\rangle
%\right]
%\end{eqnarray}
\begin{equation}
\partial_{x}u(x)=\frac{e^2 v^2_T{|\delta B_{\omega,k}|^2}}{1/{t_c}}
\frac{\Bigg(8I_1J_1-4I_2J_1+8\left(1+\frac{k^2}{4\mu^2_R}\right)I_3J_1\Bigg)}{5I_2J_2}\label{velocityderivative}
\end{equation}  
where,
\begin{eqnarray}
\nonumber
I_1=\Bigg(\frac{1}{(2\pi)^3}\int d\Omega(2v^2_{x}-v^2_y-v^2_z)v^2_x\frac{(e\delta{\bf{B_{\omega,k}\cdot v}})^2}{4\mu^4_R}=\frac{e^2\delta{\bf{B^2_{\omega,k}}}}{105\pi^2\mu^4_R}\\ \nonumber
I_2=\Bigg(\frac{1}{(2\pi)^3}\int d\Omega(2v^2_{x}-v^2_y-v^2_z)\frac{(e\delta{\bf{B_{\omega,k}\cdot v}})^2}{4\mu^4_R}-\frac{e^2\delta{\bf{B^2_{\omega,k}}}}{15\pi^2\mu^4_R},\\ \nonumber
I_3=\Bigg(\frac{1}{(2\pi)^3}\int d\Omega(2v^2_{x}-v^2_y-v^2_z)v^2_z\frac{(e\delta{\bf{B_{\omega,k}\cdot v}})^2}{4\mu^4_R}=-\frac{e^2\delta{\bf{B^2_{\omega,k}}}}{210\pi^2\mu^4_R},\\ \nonumber
J_1=\int^{\infty}_0 dp p\langle n_{\mathbf{p}}\rangle,\\ \nonumber
J_2=\int^{\infty}_0 dp p^3\langle n_{\mathbf{p}}\rangle.
\end{eqnarray}
%Here, $d_p=\frac{d}{dp}$. Only the second term contribute to the viscosity, Now taking the moment  $(2p^2_x-p^2_y-p^2_z)$ the right  
%hand side of Eq.(\ref{diffusionequation1}) we can write,
Now, using Eq.(\ref{energy-momentum}) one can write,
\begin{equation}
 (2T^{xx}-T^{yy}-T^{zz})=\int \frac{d\Omega}{(2\pi)^3}\frac{(e\delta{\bf{B_{\omega,k}\cdot v}})^2(2v^2_{x}-v^2_y-v^2_z)}{4\mu^4_R}\int^{\infty}_0{dp} 
 p^3{\langle n_{\mathbf{p}}\rangle}\label{energymomentum}
\end{equation}
The definition of shear viscosity is,
\begin{equation}
 \eta_{A}=\frac{(2T^{xx}-T^{yy}-T^{zz})}{-4\partial_{x}u(x)} \label{shearviscosity}
\end{equation}

%\begin{equation}
%\eta_{A}=\frac{{v^2_{T}}\int \frac{d\Omega}{(2\pi)^3}\frac{(e\delta{\bf{B_{\omega,k}\cdot v}})^2(2v^2_{x}-v^2_y-v^2_z)}{4\mu^3_R}\int^{\infty}_0 
%dp p^3\frac{d\langle n_{\mathbf{p}}\rangle}{dp}
%\int \frac{d\Omega}{(2\pi)^3}\frac{(e\delta{\bf{B_{\omega,k}\cdot v}})^2(2v^2_{x}-v^2_y-v^2_z)}{4\mu^3_R}\int^{\infty}_0{dp} 
% p^2{\langle n_{\mathbf{p}}\rangle}}{\frac{{e^2 v^2_T}{k^2|\delta B_{\omega,k}|^2}}{4\mu^2_R}\frac{1}{1/{t_c}}
%\Bigg(\int \frac{d\Omega}{(2\pi)^3}(2v^2_{x}-v^2_y-v^2_z)\frac{(e\delta{\bf{B_{\omega,k}\cdot v}})^2}{4\mu^3_R}\Big[\int^{\infty}_0 
%dp p\frac{d\langle n_{\mathbf{p}}\rangle}{dp}
%\Big]\Bigg)}
%\end{equation}
%\end{widetext}

%Now let us consider the case 
%%of cold plasma can be found in the interior of neutron star
%%where
%$\mu_5\sim\mu_R>>T$. In this case there will be very small contribution of anti particles. This is because in this case  distribution function for anti-particle contribution will vanish. 
Taking the distribution function of the form, 
$\langle n_{\mathbf{p}}\rangle=\frac{1}{\exp{(\mu_R-p)}+1}$, considering $\mu_R\gg T$  
and  using Eqs.(\ref{velocityderivative},\ref{energymomentum}) 
and $k=k_{max}=\frac{2\mu_R\alpha}{3\pi}$ one can estimate anomalous shear viscosity,

\begin{equation}
\eta_{A}\sim
%\frac{\mu^2_R}{t_{c}}\left(1+\frac{11\pi^2 T^2}{3\mu^2_R}\right)\sim
\frac{\mu^2_R}{t_{c}}\left(1+\frac{11\pi^2 T^2}{3\mu^2_R}\right)
\end{equation}
%Here we have replaced $\mu_R$ by $\mu_5$ because we have considered $\mu_R>>\mu_L$. 
%Now Substituting $k=k_{max}=\frac{2\mu\alpha}{3\pi}$ we get;
%
%\begin{equation}
%\eta_{A}=\frac{\pi \mu^2_5}{15 \alpha^2 t_{c}}
%\end{equation}
One can notice from Fig.(1) that that for the case $\mu_R\gg T$ and $k=k_{max}$,
$1/t_c$ depends on $\mu_R$ in an approximately linear way i.e. $1/t_c \propto \mu_R$.
%and the slope of the curve depend on $k_{max}/m_D=\frac{2\alpha}{9\pi}\ll 1$. 
Slope of the curve can be found by a linear fit. For  $\alpha=1/137$  the slope is $\sim 5.09*10^-7$. The slope increases by increasing $\alpha$. Thus $\eta$ scales like $\mu_R^3$.

\section{Conclusion}
We have calculated the coefficient of shear viscosity based on the strong turbulence
argument. For the case when $\mu_R/T \gg 1$, the collision rates becomes insignificant\cite{Chen:2012jc}
at low temperature, in this regime the decorrelation frequency $1/t_c$ can have a
significant contribution in determining $\eta$. In this  low temperature limit
the entropy density $s$ scales as $\mu_R^2T$ and the ratio $\eta/s\propto \mu_R/T$
and it can be a large number. 
%In the limit $T \rightarrow 0$, CPI
%can give finite values for the coefficient of the shear viscosity. 
In deriving the above
expression of $\eta$ we have ignored non-linear wave-wave interaction which can play
a role in case of non-Abelian plasmas. However to address this question one require to
numerically simulate the chiral plasma instability with the full nonlinearity.

Note that dimensional argument suggests that for the case
when $\mu_R\ll T$, stress (energy density) $\sim \mu^2_R
T^2$, decorrelation frequency ($1/t_c\sim\omega_{max}$)  of CPI
$\sim \mu_R$ \cite{Kumar:2016xuh} and $\eta$ scales as $\mu_R
T^2$. Therefore, $\eta/s\propto \mu_R/T$, which could be a 
small number. We hope that this analytic study will help in
understanding the viscosity due to turbulent transport
in parity violating plasma and can be useful in it numerical simulations.

\end{document}